\documentclass[twocolumn,aps,pra,showpacs]{revtex4}

\usepackage{epsfig}

\begin{document}


\title{Modified dynamics of weakly coupled BEC's Josephson Junction (BJJ) }


\author{Yu-ping Huang, Zhen-sheng Yuan, Lin-fan Zhu, Lin-jiao Luo, Xiao-jing Liu,Ke-zun Xu}
\affiliation
{ Laboratary of Bond Selective Chemistry, Department of Modern Physics,
\\University of Science and Technology of China, Hefei, Anhui 230027, China
 }


\date{\today}

\begin{abstract}
The tunnelling quantum dynamics of bimodal BJJ system is modified
through introducing an equilibrium condition, which is based on
the assumption that the BJJ is tend to keep on its ground state
(with a lowest energy) during the oscillation. The tunnelling
dynamics of BJJ with symmetric and asymmetric traps is discussed
through numerically solving the modified equations. Stationary
states are found to exist in the both BJJs. Compared to previous
works, the macroscopic quantum self trapping (MQST) is
auto-avoided. Meanwhile, it is revealed that the BJJ oscillates
with its inherent frequency which is only related to the Josephson
energy, which has been testified experimentally in other contexts.
\end{abstract}


\maketitle

\section{Introduction\label{sec-intro}}

Since Bose-Einstein condensation was first detected in 1995 \cite{anderson1995}, there have been rapid and critical
developments in experimental techniques \cite{stamper1998a, stamper1998b, hall1998a, catali2001, andrews1997}. In 1997,
interference fringes in two overlapping condensates were observed\cite{andrews1997}. Consequently in 1998, the
superposition of condensed atoms in different hyperfine levels was created \cite{stamper1998a, stamper1998b}. And the
evolution of the relative phase of two coupled condensates was measured by interferometry techniques \cite{hall1998a,
hall1998c}. In 2001, the direct observation of an oscillating atomic current in one-dimensional array Josephson
junctions was realized \cite{catali2001}. With these precise manipulation of Bose-Einstein condensates (BEC's), it has
enhanced the possibility of tailoring the new quantum systems to a degree not possible with other quantum systems, like
superfluid. And the studies of spatial coherence naturally raise the question of measurement and exploration of
temporal phase coherence between two condensates.

Theoretically, with the approximation of non-interacting atoms and small-amplitude Josephson oscillations, some aspects
of temporal phase coherence have been investigated in the context of BEC by means of Josephson junctions
\cite{javan1986, dalfovo1996}. Yet many important features of this subject remained to be explored. A non-excitation
system of interacting bosons confined within an external potential can be described by a macroscopic wave function,
which has the meaning of an order parameter and satisfies the nonlinear Gross-Pitaevskii (GP) equation
\cite{dalfovo1999}. The semiclassical tunnelling quantum dynamics using GP equation has been studied
\cite{raghava1999}. And many interesting phenomena such as macroscopic quantum self-trapping (MQST) \cite{milburn1997,
smerzi1997, raghava1999} and $\pi$ states \cite{smerzi1997, raghava1999} were predicted in a weakly coupled double BEC.
In those works, the discrete nonlinear Schr\"{o}dinger equation (DNLSE) has served as a powerful tool for describing
the boson Josephson junction (BJJ) \cite{stamper1998a, stamper1998b}. But so far, many investigations have cast doubts
on the validity of the DNLSE for conserved quasiparticles interacting with a boson field \cite{raghava1999b}. The
objections are not to the GP equation itself , but to the quasiparticle approximation instead. The earlier results at
the condition of short time scales obtained through the DNLSE in the context of coupled quasiparticle-boson systems
were verified by the fully quantum version of the BJJ tunnelling model \cite{stamper1998b, hall1998a}. But on long time
scales, the MQST in symmetric traps, which is predicted by DNLSE, was found to be destroyed in fully quantum dynamics
\cite{catali2001}.

As revealed in other context\cite{raghava1999b}, the MQST (or z-symmetry breaking) in a translationally invariant
Hamiltonian is an artifact of the factorization assumption inherent in the semiclassical dynamics. In the DNLSE
procedure, atoms are assumed to be localized in either of the traps with discrete wavefunctions. Then, the quantum
states for the total BJJ system is the superposition of those eigenstates in each trap individually. Furthermore, the
population distribution is determined through the calculation of time dependent Schr\"{o}dinger equation. Compared to
the fully quantum dynamics, in the semiclassical description, the quantum thermodynamic equilibrium for BJJ system is
overlooked. And a stationary BJJ is not restricted to its ground state with a lowest energy. A consequent result is the
artifact of MQST, which would not break down in the semiclassical description even on the condition of long time
scales.

Compared to previous works, the BJJ is assumed to occupy its ground state in present study. And a corresponding lowest
energy condition was deduced, based on which the semiclassical dynamics was modified. The numerical calculation of the
modified dynamics revealed that the MQST was auto-avoided. Also, it was found that the BJJ oscillated with an inherent
frequency, which was only related to the Josephson energy.

In the following Sec. \ref{sec-bjj}, the equilibrium behavior of BJJ is discussed, and the lowest energy state is found
under a condition of a fixed relationship between the population distribution and the quantum states of BEC's. In Sec.
\ref{sec-taunel}, the semiclassical dynamics is modified by introducing the stationary condition into the DNLSE
procedure. In Sec. \ref{sec-traps}, the tunnelling dynamics for BJJ with symmetric and asymmetric traps is discussed
and some numerical results are given.

\section{The equilibrium behavior of weakly coupled BJJ \label{sec-bjj}}

The effective many-body Hamiltonian describing atomic BEC in a
double-well trapping potential $V_{trap}(r)$ can be written in the
second-quantization form as \cite{chen2002}

\begin{eqnarray} \label{eqn-ham}
\hat{H}=\int d^3 r \hat{\Psi}^{\dagger}[-\frac{\hbar^2}{2m}
\nabla^2 + V_{trap}] \hat{\Psi} + \frac{U_0}{2} \int d^3 r
\hat{\Psi}^{\dagger} \hat{\Psi}^{\dagger} \hat{\Psi} \hat{\Psi}.
\end{eqnarray}

\noindent Here $m$ is the atomic mass and $U_0 = 4 \pi a \hbar^2
/m$ ($a$ denotes the s-wave scattering length, measuring the
strength of the two-body atomic interaction). The Heisenberg
atomic field operators $\hat{\Psi}^{\dagger}$ and $\hat{\Psi}$
satisfy the standard bosonic commutation relation
$[\hat{\Psi}^{\dagger}(r,t), \hat{\Psi}(r^\prime,t)]=\delta
(r-r^\prime)$. In the quasiparticle approximation, one can expand
the field operators $\hat{\Psi}$ in terms of two local modes

\begin{eqnarray} \label{eqn-psi}
\hat{\Psi}(r,t)\approx \sum_{i=1,2} \hat{a}_i(t) \psi_i(r).
\end{eqnarray}

\noindent Here $[\hat{a}_i(t), \hat{a}_i^{\dagger}(t)] =
\delta_{ij}$, and $\psi_i(r)$ stands for the local mode function
of either well and satisfies

\begin{eqnarray} \label{eqn-psi2}
\int d^3 r \psi_i^*(r) \psi_i (r) \approx \delta_{ij}.
\end{eqnarray}

\noindent Substituting Eq. (\ref{eqn-psi}) into Hamiltonian
(\ref{eqn-ham}), the two-mode approximation of $H$ is yielded

\begin{eqnarray} \label{eqn-ham2}
\hat{H}=\sum_{i=1,2} (E_i^0 \hat{a}_i^{\dagger} \hat{a}_i +
\lambda_i \hat{a}_i^{\dagger} \hat{a}_i^{\dagger} \hat{a}_i
\hat{a}_i) - (J \hat{a}_1^{\dagger} \hat{a}_2 + J^* \hat{a}_1
\hat{a}_2^{\dagger}),
\end{eqnarray}

\noindent where the parameters are estimated by \cite{milburn1997}

\begin{eqnarray} \label{eqn-earray}
E_i^0 = \int d^3 r \psi^* [-\frac{\hbar^2}{2m}\nabla^2 + V_{trap}]
\psi_i,  \nonumber \\  \lambda_i=\frac{U_0}{2} \int d^3 r
|\psi_i|^4, \\ J=-\int d^3 r [\frac{\hbar^2}{2m}\nabla\psi_1^*
\cdot \nabla\psi_2 + V_{trap} \psi_1^* \psi_2].  \nonumber
\end{eqnarray}

\noindent Here the interactions between atoms in different wells
are neglected as in the weakly coupled BEC's. Defining two local
number operators $\hat{n}_i=\hat{a}_i^{\dagger} \hat{a}_i$, it is
easy to verify that the total number operator
$\hat{N}=\hat{a}_1^{\dagger} \hat{a}_1 + \hat{a}_2^{\dagger}
\hat{a}_2=N_T$ represents a conserved quantity. After neglecting
the constant term, the two-mode Hamiltonian $\hat{H}$ can be
rewritten as\cite{chen2002}:

\begin{eqnarray} \label{eqn-harray}
\hat{H}^{\prime} = E_c(\hat{n}-\hat{n}_g)^2-(J\hat{a}_1^*
\hat{a}_2 + J^* \hat{a}_1 \hat{a}_2^*),  \nonumber \\
\hat{n}=(\hat{n}_1-\hat{n}_2)/2, \nonumber \\ E_c=\lambda_1 +
\lambda_2, \\ \hat{n}_g = \frac{1}{2E_c}[(E_2^0-E_1^0)+(N_T-1)
(\lambda_1-\lambda_2)]. \nonumber
\end{eqnarray}

\noindent Here $\hat{n}$ is the number difference operator, $E_c$
is the ``charging energy'' and $\hat{n}_g$ is known as the ``gate
charge'' \cite{makhlin2001}.

Generally, the BJJ system woks in Fock regime, which gives $E_c
\gg J$, and Hamiltonian (\ref{eqn-harray}) can be well
approximated by

\begin{eqnarray} \label{eqn-happrox}
\hat{H}^{\prime}=E_c(\hat{n}-\hat{n}_g)^2.
\end{eqnarray}

\noindent A BJJ system in ground state should yield a minimized
energy (or say a maximized entropy) and the Hamiltonian
(\ref{eqn-happrox}) should be approximate zero, which gives

\begin{eqnarray} \label{eqn-number}
\hat{n}=\hat{n}_g.
\end{eqnarray}

\noindent Eq. (\ref{eqn-number}) denotes the relationship between the population distribution and ``gate charge'' of an
stationary BJJ system in its energy ground state. It is important to notice that the ``gate charge'', which is related
to the eigenstates of BEC in the either trap, is determined by two components, i.e. the boson-field interaction and
self interaction, as shown in Eq. (\ref{eqn-harray}). In the present work, the number difference $\hat{n}=\hat{n}_g$ is
not necessarily an integer, and the artifact of the factorization of assumption inherent in the semiclassical dynamics
is automatically avoided. In this work, the boson-field interaction energy is considered as independent of the number
difference $n$, while the self interaction energy is shifted by the atom numbers in either trap \cite{dalfovo1999}.

\section{Modified tunnelling dynamics \label{sec-taunel}}

In the DNLSE procedure, the wavefunction Eq. (\ref{eqn-psi}) is
rewritten in the quasiparticle form, with

\begin{eqnarray} \label{eqn-wav}
\Psi(r,t) \approx \sum_{i=1,2} \phi_i(t) \psi_i(r),  \nonumber \\
\phi_{1,2}(t)=\sqrt{N_{1,2}(t)} e^{i\theta_{1,2}(t)},
\end{eqnarray}

\noindent  where $N$ and $\theta$ denote the population and phase
shift of BJJ, and $\phi_i$ represents the local mode function
$\hat{a}_i(t)$. Then, the Hamiltonian (\ref{eqn-harray}) is
reformed to

\begin{eqnarray} \label{eqn-wavarray}
i\hbar \frac{\partial \phi_1}{\partial t}=(E_1^0 + \lambda_1
N_1)\phi_1 - J \phi_2 ,    \nonumber \\
i\hbar \frac{\partial \phi_2}{\partial t}=(E_2^0 + \lambda_2
N_2)\phi_2 - J \phi_1.
\end{eqnarray}

Defining the fractional population imbalance and relative phase
\cite{andrews1997} as

\begin{eqnarray} \label{eqn-zbeta}
z(t)=(N_1(t)-N_2(t))/N_T,  \nonumber  \\
\beta(t)=\theta_1(t)-\theta_2(t).
\end{eqnarray}

\noindent Eq.(10) becomes

\begin{eqnarray} \label{eqn-zbetap}
\dot{z}(t)=-\sqrt{1-z(t)^2}\sin [\beta(t)], \nonumber  \\
\dot{\beta}(t)=\Delta E /(2J) + \frac{z(t)}{\sqrt{1-z(t)^2}} \cos
[\beta(t)].
\end{eqnarray}

\noindent Here $t$ has been rescaled to a dimensionless time $t2J/h$. And $\Delta E$ denotes the particle energy
difference between atoms in either traps, which acts as an asymmetric parameter in the tunnelling quantum dynamic of
BJJ. In present work, the BJJ system is assumed to stay at its ground state. So the stationary condition Eq.
(\ref{eqn-number}) should be met. By introducing Eq. (\ref{eqn-number}), Eq. (\ref{eqn-zbetap}) should be modified. It
was found that the dynamic equations keep its form as shown by Eq. (\ref{eqn-zbetap}), but with the energy difference
modified to

\begin{eqnarray} \label{eqn-energy}
\Delta E =(E_1^0 -E_2^0)/2.
\end{eqnarray}

\noindent And the total conserved energy is given by

\begin{eqnarray} \label{eqn-hnew}
H=-\frac{1}{2J} \Delta E \cdot z -\sqrt{1-z^2} \cos \beta.
\end{eqnarray}

Eq. (\ref{eqn-zbetap}) can be rewritten in the Hamiltonian form
\cite{raghava1999}

\begin{eqnarray} \label{eqn-zbetapp}
\frac{\partial z}{\partial t} = - \frac{\partial H}{\partial
\beta}, \nonumber \\
\frac{\partial \beta}{\partial t} = \frac{\partial H}{\partial z}.
\end{eqnarray}

In deducing Eq. (\ref{eqn-zbetapp}), we have put forward the fact that at low oscillation frequency, the BJJ system
would stay in its lowest energy state, say ground state, and the population distribution would be related to the ``gate
charge''. In the previous tunnelling quantum dynamic description of BJJ, the GP equation describing the mean-field
dynamics of a BEC is reformed to the bimodal discrete nonlinear Schr\"{o}dinger equation, and the calculation is
carried out by solving self-consistent equation in each trap respectively \cite{willian1999}. It would bring about
artifact since the equilibrium behavior has been overlooked. And a consequent result of previous tunnelling dynamics is
the MQST \cite{raghava1999b}. Generally, MQST is based on the condition that the initial conserved energy

\begin{eqnarray} \label{eqn-mqstcon}
H_0=H[z(0),\beta(0)]>1,
\end{eqnarray}

\noindent where $z(0)$ and $\beta(0)$ are the initial values of
$z$ and $\beta$ respectively. In the earlier work
\cite{raghava1999}, the energy gap $\Delta E$ is the function of
population imbalance $z$ , which would result in some certain
initial value of $z$ satisfying Eq. (\ref{eqn-mqstcon}). Thus the
BJJ system would maintain in MQST state even at the condition of
long time scales. Compared to the previous work, a equilibrium
condition Eq. (\ref{eqn-number}) was introduced to modify DNLSE
procedure in present work. It was found the artifact induced by
DNLSE is auto-avoided because in the BJJ with symmetric traps the
expression $H_0 = -\sqrt{1-z} \cos \beta \leq 1$ is satisfied by
any given $z(0)$ and $\beta(0)$. Naturally, because the tunnelling
quantum dynamics behavior is restricted by equilibrium condition,
it is a self-consisted result of Eq. (\ref{eqn-number}) and Eq.
(\ref{eqn-zbetap}) that the oscillation frequency should be
independent of initial condition $z(0)$ and $\beta(0)$.

\section{Tunneling dynamics of BJJ with symmetric and asymmetric traps \label{sec-traps}}

\subsection{The symmetric trap case \label{sec-trap-s}}

For a symmetric BEC Josephson Junction (BJJ), the motion equations (\ref{eqn-zbetap}) is given by

\begin{eqnarray} \label{eqn-zbetasym}
\dot{z}(t)=-\sqrt{1-z(t)^2}\sin[\beta(t)], \nonumber \\
\dot{\beta}(t)=\frac{z(t)}{\sqrt{1-z(t)^2}} \cos [\beta(t)],
\end{eqnarray}

\noindent with a conserved energy $H=-\sqrt{1-z^2}\cos \beta$. The ground state is a symmetric stationary solution of
Eq. (\ref{eqn-zbetasym}), with a conserved energy and $E_- = -1$ and

\begin{eqnarray} \label{eqn-solusym}
\beta=2n\pi, \nonumber \\
z=0.
\end{eqnarray}

The next stationary state with higher energy $E_+=1$  is an
antisymmetric solution with

\begin{eqnarray} \label{eqn-soluasym}
\beta=(2n+1)\pi, \nonumber \\
z=0.
\end{eqnarray}

\begin{figure}[htp]
\epsfig{figure=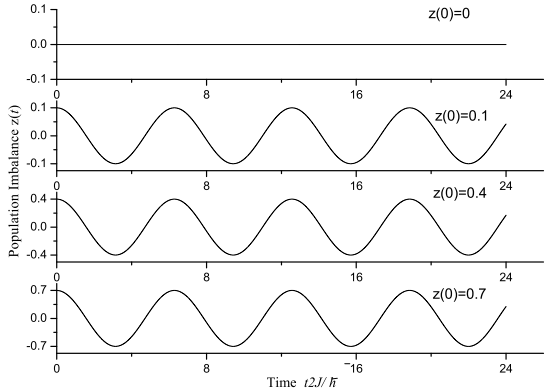, width=8.0cm} \caption{Population imbalance as a function of dimensionless time $t2J/\hbar$.
The BJJ is of symmetric traps with initial relative phase $\beta(0)=0$, and the initial population imbalance $z(0)$
takes the values 0, 0.1, 0.4 and 0.7 respectively. \label{fig-sym}}
\end{figure}

The imbalanced initial population, i.e. $z(0)\neq 0$, would result
in an oscillating solution of Eq. (\ref{eqn-zbetasym}), as shown
in Fig. \ref{fig-sym}. It is seen that for a given $z(0)$, the
population imbalance oscillates around it's zero-value, and the
z-symmetry of bi-mode BJJ is preserved. Meanwhile, the oscillation
amplitude $A$ is determined by the initial population imbalance
$z(0)$, i.e. $A=|z(0)|$.

\subsection{The asymmetric trap case \label{sec-trap-a}}

\begin{figure}[hbp]
\epsfig{figure=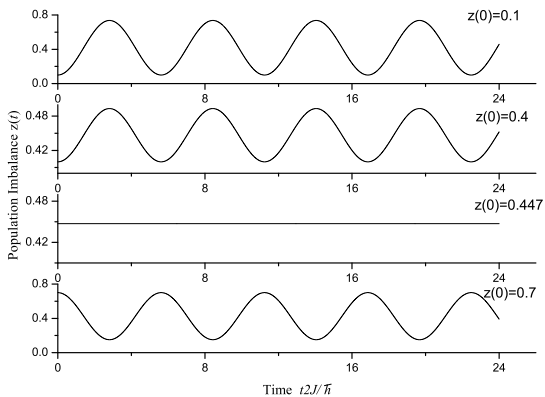, width=8.0cm} \caption{Population imbalance as a function of dimensionless time $t2J/\hbar$.
The BJJ is of asymmetric traps with $\Delta E /2J=-0.5$, and a initial relative phase $\beta(0)=0$, with the initial
population imbalance $z(0)$ takes the value $0.1$, $0.4$, $0.447$and $0.7$ respectively. \label{fig-asym}}
\end{figure}

\begin{figure}[htp]
\epsfig{figure=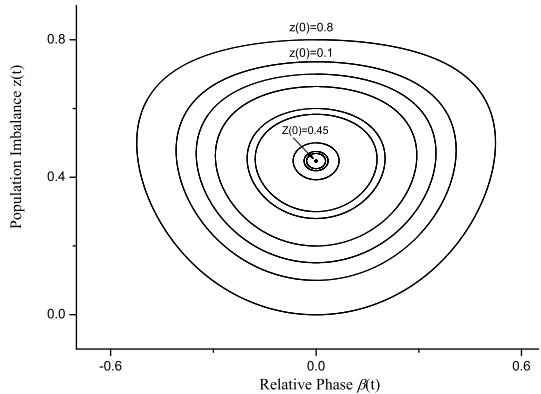, width=8.0cm} \caption{Population
imbalance as a function of relative phase $\beta(t)$. The BJJ is
of asymmetric traps with $\Delta E /2J=-0.5$, and a initial
relative phase $\beta(0)=0$. As the initial population imbalance
$z(0)$ approach a mean value of about $\bar{z}=0.45$, the
oscillating amplitude approach zero.\label{fig-3}}
\end{figure}

When the trap is asymmetric, the energy difference is not equal to
zero, i.e. $\Delta E \neq 0$. The numerical solution of Eq.
(\ref{eqn-zbetap}) is given in Fig.2 with the trap asymmetry
number $\Delta E/2J =1$ and $z(0)=0$, $0.4$, and $0.7$
respectively. It is shown that the population imbalance $z(t)$
oscillates around its mean value $\bar{z}$ with an amplitude
$A=|z(0)-\bar{z}|$. Also the $z(t)$ plotted against relative phase
$\beta(t)$ is given in Fig.3. It is expected that when
$z(0)=\bar{z}=0.447$, oscillating amplitude would be zero, i.e.
$A=0$, and the BJJ system should arrive at a stationary state. The
corresponding ground state is obtained with the symmetric phases,

\begin{eqnarray} \label{eqn-soluaasm1}
\beta=2n\pi,  \nonumber \\
z=\bar{z}.
\end{eqnarray}

The next stationary state with higher conserved energy is given by

\begin{eqnarray} \label{eqn-soluaasm2}
\beta=(2n+1) \pi,  \nonumber \\
z=\bar{z},
\end{eqnarray}

with antisymmetric phases. Moreover, as shown in Eq.
(\ref{eqn-zbetap}), the stationary state may be arrived only on
the condition of $z(0)=\bar{z}$, and is independent of its initial
relative phases.

It is important to notice that, both in Fig. \ref{fig-sym} and
Fig. \ref{fig-asym}, the oscillation frequency $f$ is independent
of either the trap asymmetry number $\Delta E/2J$ or the initial
value $z(0)$. It is the inherent character of BJJ, which is only
related to the Josephson energy $J$, with

\begin{eqnarray}\label{eqn-frequence}
f \propto \frac{\hbar}{J}
\end{eqnarray}

\noindent Since Josephson energy is determined by the interwell
potential height, it is experimentally feasible to testify the
modified dynamics through Eq.\ref{eqn-frequence}. In fact,it has
provided a testification of the present
scheme\cite{cataliotti2001}.

\section{Conclusions}

In conclusion, the tunnelling quantum dynamics of bimodal BJJ system is modified under an assumption that the BEC's is
tend to keep on the ground state with a lowest energy. And the lowest energy is determined by the population
distribution. The dynamics of BJJ with symmetric and asymmetric traps is discussed with the modified equations.
Compared with previous works, the MQST is auto-avoided in this work, as has been testified by previous experimental
works. Also, it is revealed that the BJJ oscillates with its inherent frequency, which is only determined by the
Josephson energy. This has been testified experimentally. To the end, it is necessary to point out that it remains to
be seen how quasipaticle excitation and energy dissipation of BEC would affect the tunnelling dynamics.

\begin{acknowledgments}

The discussion with Prof. Zeng-bing Chen is very helpful to this
work. And supports by National Nature Science Fund of China
(10134010) and the Youth Foundation of the University of Science
and Technology of China are gratefully acknowledged.

\end{acknowledgments}

\end{document}